\documentstyle[prl,aps,graphics,epsfig]{revtex}  
\input psfig.sty
\begin{document}

\title{Large Momentum Transfer Measurements of the Deuteron Elastic 
Structure Function ${\mathbf A(Q^2)}$ at Jefferson Laboratory}

\author {
L.C.~Alexa,$^{25}$
B.D.~Anderson,$^{14}$
K.A.~Aniol,$^{2}$
K.~Arundell,$^{34}$
L.~Auerbach,$^{30}$
F.T.~Baker,$^{7}$
J.~Berthot,$^{1}$
P.Y.~Bertin,$^{1}$
W.~Bertozzi,$^{18}$
L.~Bimbot,$^{23}$
% ...
and 
Z.-L.~Zhou.$^{18}$
\vspace{.2cm}
} 
\address{ 
{ (The Jefferson Lab Hall A Collaboration) } \\
\vspace{.1cm}
$^{1}$Universit\'{e} Blaise Pascal/IN2P3, F-63177 Aubi\`{e}re, France. 
$^{2}$California State University at Los Angeles, Los Angeles, CA 90032.
% ...
$^{7}$University of Georgia, Athens, GA 30602.
% ...
$^{14}$Kent State University, Kent OH 44242.
% ...
$^{18}$Massachusetts Institute of Technology, Cambridge, MA 02139.
% ...
$^{23}$Institut de Physique Nucl\'{e}aire, F-91406 Orsay, France.
% ...
$^{25}$University of Regina, Regina, SK S4S OA2, Canada.
% ...
$^{30}$Temple University, Philadelphia, PA 19122.
% ...
$^{34}$College of William and Mary, Williamsburg, VA 23187.
% ...
\vspace{0.2cm}\\
{\rm Submitted to Physical Review Letters}
}

\date{October 28, 1998}
\maketitle

\begin{abstract}

The deuteron elastic structure function $A(Q^2)$ has been
extracted in the range $0.7 \le Q^2 \le 6.0$ (GeV/c)$^2$
from cross section measurements of elastic electron-deuteron scattering
in coincidence using the Hall A Facility of Jefferson 
Laboratory.  The data are compared to theoretical models based on 
the impulse approximation with the inclusion of meson-exchange 
currents, and to predictions of quark dimensional scaling and 
perturbative quantum chromodynamics. 

\end{abstract}


\twocolumn
\narrowtext

Electron scattering from the deuteron has long been a
crucial tool in understanding the internal structure and 
dynamics of the nuclear two-body system.  In particular,
the deuteron electromagnetic form factors, measured in elastic scattering, 
offer unique opportunities to test models of the short-range nucleon-nucleon 
interaction, meson-exchange currents and isobaric configurations as well 
as the possible influence of explicit quark degrees of 
freedom\cite{CA98},\cite{CA97}. 

The cross section for elastic electron-deuteron~(e-d) scattering
is described by the Rosenbluth formula:
\begin{equation} 
{{d\sigma} \over {d\Omega}} = \sigma_M \left[ A(Q^2) +
B(Q^2) \tan^2{(\frac{\theta}{2})} \right]  
\end{equation}
where $\sigma_M=\alpha^2 E^\prime \cos^2(\theta/2)/[4 E^3 \sin^4(\theta/2)]$
is the Mott cross section.  
Here $E$ and $E^\prime$ are the incident and scattered electron
energies, $\theta$ is the electron scattering angle,
$Q^2=4 E E^\prime \sin^2(\theta/2)$ is the four-momentum transfer
squared and $\alpha$ is the fine structure constant.  The elastic
electric and magnetic structure functions $A(Q^2)$ and $B(Q^2)$ are
given in terms of the charge, quadrupole and magnetic form factors
$F_{C}(Q^2)$, $F_{Q}(Q^2)$, $F_{M}(Q^2)$:
\begin{eqnarray}
{ A(Q^2) =  { F^2_{C}(Q^2) + {8 \over 9} \tau^2 F^2_{Q}(Q^2) +
{2 \over 3} \tau F^2_M(Q^2) }  } \\
{ B(Q^2) =  {4 \over 3} \tau (1+\tau) F^2_M(Q^2) }
\end{eqnarray}
where $\tau=Q^2/4M^2_d$, with $M_d$ being the deuteron mass.
 
The interaction between the ~electron and the deuteron is mediated
by the exchange of a virtual ~photon.  In the non-relativistic impulse 
approximation (IA), where the ~photon is assumed to ~interact 
with one of the ~two nucleons in the ~deuteron,
the deuteron ~form factors are described in ~terms of the deuteron wave
function and the ~electromagnetic form factors ~of the nucleons.
Theoretical calculations based on ~the IA approach\cite{CA98} using various
nucleon-nucleon ~potentials and parametrizations of the ~nucleon
form factors ~generally underestimate the ~existing $A(Q^2)$ 
data \cite{EL69},\cite{AR75},\cite{CR85},\cite{PL90}.
Recent relativistic ~impulse approximation (RIA) calculations ~improve 
or worsen the agreement ~with the data depending ~on their particular 
assumptions.  There are ~two RIA approaches: manifestly ~covariant 
calculations \cite{OR95},\cite{HU90} ~and 
light-front dynamics\cite{CH88},\cite{JC98}.
 
It is well known that the form factors of the deuteron
are very sensitive to the presence of meson-exchange currents 
(MEC)\cite{CA98}.
There have been numerous extensive studies and calculations
augmenting both the IA and RIA approaches with the inclusion of MEC.
The principal uncertainties in these calculations are the
poorly known value of the $\rho\pi\gamma$ coupling constant and
the $\rho\pi\gamma$ vertex form factor.
Some calculations show also sensitivity to possible presence
of isobar configurations in the deuteron\cite{DY90}.  
The inclusion of MEC brings the theory into better agreement
with the existing data. 
 
It is widely recognized that the underlying quark-gluon dynamics 
cannot be ignored at distances much less than the
nucleon size.  This has led to the formulation of so-called
hybrid quark models\cite{CH87}
that try to simultaneously incorporate the
quark- and gluon-exchange mechanism at short distances and the
meson-exchange mechanism at long and intermediate distances.
When the internucleon
separation is smaller than $\sim1$ fm, the deuteron is treated
as a six-quark configuration with a certain probability that results
in an additional contribution to the deuteron form factors.

At sufficiently large momentum transfers the form factors are expected
to be calculable in terms of only quarks and gluons within the framework
of quantum chromodynamics (QCD).  
The first attempt at a quark-gluon description of the deuteron form 
factors was based on quark-dimensional scaling (QDS)\cite{BR73}.
The underlying dynamical mechanism during e-d scattering 
is the rescattering of the constituent quarks via the exchange of
hard gluons.  The $Q^2$ dependence of this process is then predicted by
simply counting the number of gluon propagators (5), which implies that
$\sqrt{A(Q^2)} \sim (Q^2)^{-5}$.
This prediction was later substantiated in the framework of perturbative QCD
(pQCD), where it was shown\cite{BR83} that in leading-order:
$\sqrt{A(Q^2)} = [\alpha_{s}(Q^{2})/Q^{2}]^{5}
\sum_{m,n}d_{mn}[\ln(Q^{2}/\Lambda^{2})]^{-\gamma_{n} -\gamma_{m}}$
where $\alpha_s(Q^2)$ and $\Lambda$ are the QCD 
strong coupling constant and scale parameter, and $\gamma_{m,n}$ and $d_{mn}$
are QCD anomalous dimensions and constants.
The existing SLAC $A(Q^2)$ data\cite{AR75} exhibit some evidence of this 
asymptotic fall-off for $Q^2>2$ (GeV/c)$^2$.  

The unique features of the Continuous Electron Beam Accelerator
and Hall A Facilities of the Jefferson Laboratory (JLab) offered 
the opportunity to extend the 
kinematical range of $A(Q^2)$ and to resolve inconsistencies in
previous data sets from different laboratories by measuring
the elastic e-d cross section for
$0.7 \le Q^2 \le 6.0$ (GeV/c)$^2$.
Electron beams of 100$\%$ duty factor were scattered off a 
liquid deuterium target in Hall A.  Scattered electrons were detected
in the electron High Resolution Spectrometer (HRSE).  To
suppress backgrounds and separate elastic from inelastic processes, 
recoil deuterons were detected in coincidence with 
the scattered electrons in the hadron HRS (HRSH).  Elastic electron-proton 
(e-p) scattering in coincidence was used to calibrate this double-arm
system.  A schematic of the Hall A
Facility as used in this experiment is shown in Figure 1.

The incident beam energy was varied between 3.2 and 4.4 GeV.
The beam intensity, 5 to 120 $\mu$A, was monitored using two resonant 
cavity beam current monitors (BCM) upstream of the target system.  The two
cavities were frequently calibrated against a
parametric current transformer monitor (Unser monitor)\cite{UN91}.
The beam was rastered on the target in both horizontal and vertical 
directions at high frequency and its position was monitored with
two beam position monitors (BPM).  
The uncertainties in the incident beam current and energy were estimated 
to be $\pm 2\%$ and $\pm 0.2\%$, respectively.  

The target system contained liquid hydrogen and deuterium
cells of length $T$=15~cm.  Two Al foils separated by 15~cm 
were used to
measure the contribution to the cross section from the
Al end-caps of the target cells.  The liquid hydrogen(deuterium)
was pressurized to 1.8(1.5)~atm and pumped at high velocity ($\sim0.5$~m/s)
through the cells to heat exchangers.  The hydrogen(deuterium) 
temperature was 19(22)~K.  This system
provided a record high luminosity of $4.0 \times 10^{38}$cm$^{-2}$s$^{-1}$
($4.7 \times 10^{38}$cm$^{-2}$s$^{-1}$) for hydrogen(deuterium).  
The raster system kept beam-induced
density changes at a tolerable level: up to 2.5\%(5.0\%) at 120$\mu$A for
deuterium(hydrogen).

\vskip-0.5in
\begin{figure}[h]
\begin{center}
\psfig{file=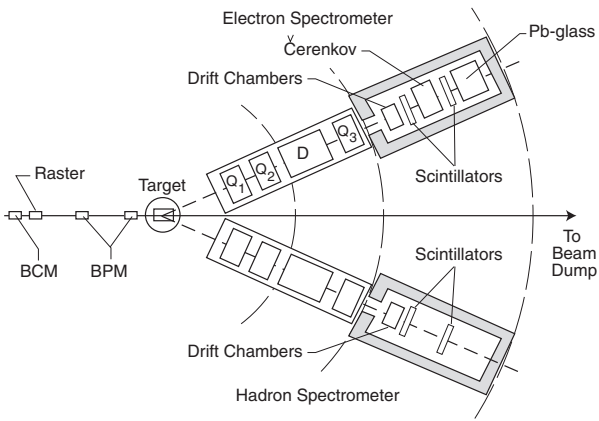,width=3.2in,angle=90}
\vskip-0.5in
\caption[]{Plan view of the Hall A Facility of JLab as used in this
experiment.  Shown are the beam monitoring devices, the cryo-target, the
two magnetically identical spectrometers (consisting of quadrupoles
Q$_1$, Q$_2$, Q$_3$, and dipole D), and the detector packages.}
\end{center}
\end{figure}    

\vspace{-0.3cm}
Scattered electrons were detected in HRSE used in its
standard configuration consisting of two planes of plastic
scintillators to form an ``electron'' trigger, a pair of drift 
chambers for electron track reconstruction, and a gas threshold 
\v{C}erenkov counter and a segmented lead-glass calorimeter for electron
identification.  Recoil nuclei were detected in HRSH
using a subset of its detection system: two planes
of scintillators to form a ``recoil'' trigger
and a pair of drift chambers for recoil track reconstruction.
The efficiencies of the calorimeter and \v{C}erenkov counter were $\sim99.5\%$,
and of scintillators and tracking almost 100$\%$ for both spectrometers. 
Event triggers consisted of electron-recoil coincidences and
of a prescaled sample of electron and recoil single-arm triggers.  

Electron events were identified on the basis of a minimal 
pulse height in the \v{C}erenkov counter and an energy deposited
in the calorimeter consistent with the momentum determined
from the drift chamber track.  Coincidence events were
identified using the relative time-of-flight (TOF) between
the electron and recoil triggers.   
Contributions from the target cell end-caps and random
coincidences were negligible.
Elastic e-p scattering was measured for each e-d elastic kinematics.
The e-p kinematics was chosen to match the electron-recoil solid angle
Jacobian for the corresponding e-d kinematics.  Data were taken with and
without acceptance defining collimators in front of the
spectrometers.  

The elastic e-p and e-d cross sections were calculated using:
\begin{equation}
\frac {d\sigma} {d\Omega}=\frac {N_{ep(d)}C_{eff}}
{N_i N_t (\Delta\Omega)_{MC} F}
\end{equation}
where $N_{ep(d)}$ is the number of e-p(e-d) elastic events,
$N_i$ is the number of incident electrons, $N_t$ is the
number of target nuclei/cm$^2$, $(\Delta\Omega)_{MC}$ is the
effective double-arm acceptance from a Monte Carlo simulation,
$F$ is the portion of
radiative corrections that depends only on $Q^2$ and $T$ (1.088 and
1.092, on average, for e-p and e-d elastic respectively)
and $C_{eff}=C_{det}C_{cdt}C_{rni}$.  Here $C_{det}$ is the
electron and recoil detector and trigger inefficiency correction (2.6$\%$),
$C_{cdt}$ is the computer dead-time correction (typically 
10$\%$ for e-d elastic), and $C_{rni}$
is the correction for losses of recoil nuclei due to
nuclear interactions in the target 
(0.7-1.8$\%$ for protons and 2.8-5.1$\%$ for deuterons).  

\begin{figure}[h]
\begin{center}
\psfig{file=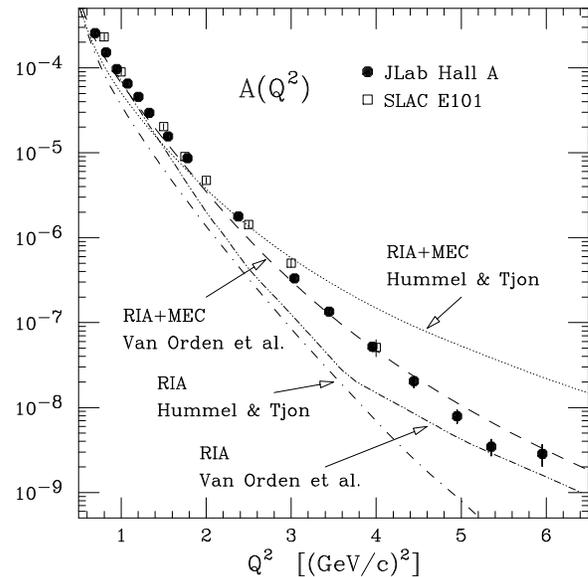,angle=90,width=3.0in}
\caption[]{The deuteron elastic structure function $A(Q^2)$ from this 
experiment compared to RIA theoretical calculations\cite{OR95},\cite{HU90}.
Also shown are previous SLAC data\cite{AR75}.}
\end{center}
\end{figure} 

\vspace{-0.3cm}
The effective double-arm acceptance was evaluated with a
Monte Carlo computer program that simulated elastic
e-p and e-d scattering under identical conditions as our
measurements.  The program tracked scattered electrons and
recoil nuclei from the target to the detectors through the 
two HRS's using optical models based on magnetic measurements
of the quadrupole and dipole elements and on position surveys
of collimation systems, magnets and vacuum apertures.  The
effects from ionization energy losses and multiple scattering
in the target and vacuum windows were taken into account for both
electrons and recoil nuclei.  
Bremsstrahlung radiation losses 
for both incident and scattered electrons in the target and
vacuum windows as well as internal radiative effects were also
taken into account.  Details on this simulation method can
be found in Ref. \cite{KA86}.
Monte Carlo simulated spectra of scattered electrons and recoil 
nuclei were found to be in very good agreement with experimentally
measured spectra.

\begin{figure}[h]
\begin{center}
\psfig{file=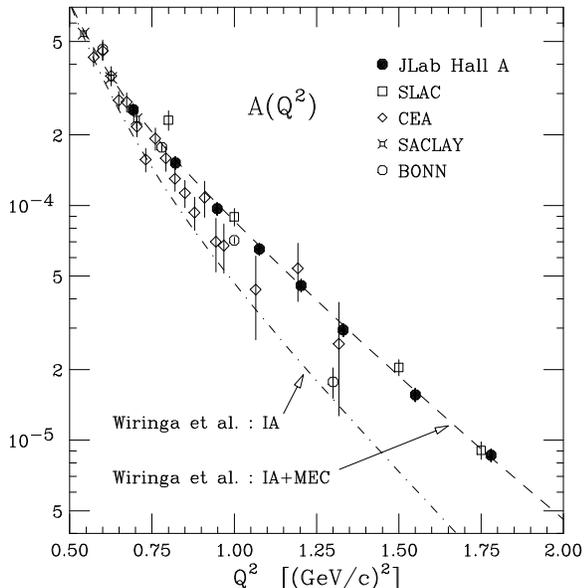,angle=90,width=3.0in}
\caption[]{The present $A(Q^2)$ data compared with overlapping data from
CEA\cite{EL69}, SLAC\cite{AR75}, Bonn\cite{CR85}, Saclay\cite{PL90},
and IA+MEC theoretical calculations\cite{WI95}.}
\end{center}
\end{figure} 

\vspace{-0.3cm}
The e-p elastic cross sections measured with the acceptance defining
collimators were found to agree within 
0.3$\%$, on average, with values calculated 
using a recent fit\cite{BO95} to world data of the proton form factors.  
The e-p elastic cross sections measured without the collimators were,
on average, 2.6$\%$ higher than the ones measured with collimators. 
All e-d cross section data taken without collimators have been normalized 
by 2.6$\%$.

Values for $A(Q^2)$ were extracted from the measured e-d cross
sections under the assumption that $B(Q^2)$ does not contribute
to the cross section (supported by the 
existing $B(Q^2)$ data\cite{AR87}).  The extracted $A(Q^2)$
values are presented in Fig. 2 together with previous SLAC 
data\cite{AR75} and theoretical calculations.  
The error bars represent statistical and systematic uncertainties
added in quadrature.  The statistical error ranged from 
$\pm1\%$ to $\pm28\%$.  The systematic error has been
estimated to be $\pm5.9\%$ and is dominated by the uncertainty
in $(\Delta\Omega)_{MC}$ ($\pm3.6\%$).
Each of the two highest $Q^2$ points represents the average of two 
measurements with different beam energies (4.0 and 4.4 GeV).
Tables of numbers are given in Ref.~\cite{WE98}.
It is apparent that our data agree
very well with the SLAC data in the range of overlap and exhibit a 
smooth fall-off with $Q^2$.  

The double dot-dashed and dot-dashed curves in Fig. 2 represent the
RIA calculations of Van Orden, Devine and Gross (VDG) \cite{OR95} and Hummel 
and Tjon (HT)\cite{HU90}, respectively.  The VDG curve is based on a 
relativistically covariant calculation that uses the Gross 
equation\cite{GR69} and assumes that the virtual photon
is absorbed by an off-mass-shell nucleon or a nucleon that is
on-mass-shell right before or after the interaction.
The HT curve is based on a one-boson-exchange quasipotential approximation 
of the Bethe-Salpeter equation\cite{SA51} where the two nucleons are treated 
symmetrically by putting them equally off their mass-shell with 
zero relative energy.  In both cases the RIA appears to be lower 
than the data.  Both groups have augmented their models by including
the $\rho\pi\gamma$ MEC contribution.  The magnitude of 
this contribution depends on the $\rho\pi\gamma$ coupling constant
and vertex form factor choices\cite{IT93}.  The VDG model (dashed curve) 
uses a $\rho\pi\gamma$ form factor from a covariant separable quark
model\cite{MI95}.  The HT model (dotted curve) uses a Vector
Dominance Model.  The difference in the two models is indicative of
the size of theoretical uncertainties.  Although our data favor the 
VDG calculations, a complete test of the RIA+MEC framework will
require improved and/or extended measurements of the nucleon form
factors and of the deuteron $B(Q^2)$, planned at JLab. 

\begin{figure}[h]
\begin{center}
\psfig{file=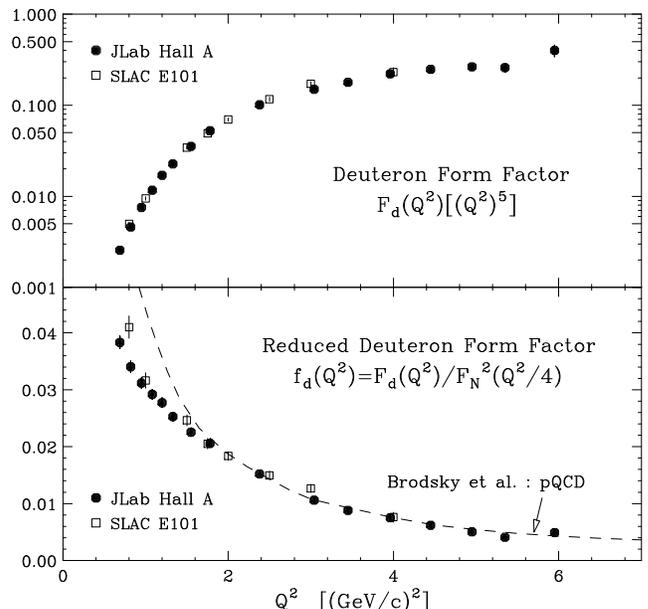,angle=90,width=3.3in}
\caption[]{The deuteron form factor $F_d(Q^2)$ times $(Q^2)^5$ (top)
and the reduced deuteron form factor $f_d(Q^2)$ (bottom) from this experiment
and from SLAC\cite{AR75}.  The curve is the asymptotic
pQCD prediction of Ref. \cite{BR83} for $\Lambda$=100~MeV, 
arbitrarily normalized to the data at $Q^2=4$~(GeV/c)$^2$.}
\end{center}
\end{figure}

\vspace{-0.3cm}
Figure 3 shows our data in the ``low'' $Q^2$ range where they
overlap with data from other laboratories.  The previous measurements
tend to show two long-standing diverging trends, one supported by 
the SLAC data and the other one by the CEA\cite{EL69} and
Bonn\cite{CR85} data.  Our data agree with the Saclay data\cite{PL90} and
confirm the trend of the SLAC data.  It should be noted that another 
JLab experiment has measured $A(Q^2)$ in the $Q^2$ range 0.7 to 1.8
(GeV/c)$^2$\cite{AB98}.
The two curves are from a recent non-relativistic IA 
calculation\cite{WI95} using the
Argonne $v_{18}$ potential without (dot-dashed curve) and with (dashed curve)
MEC, and exhibit clearly the necessity of MEC inclusion also in the 
non-relativistic IA.   

Figure 4 (top) shows values for the ``deuteron form factor'' 
$F_d(Q^2) \equiv \sqrt{A(Q^2)}$ multiplied by $(Q^2)^5$.  It is evident that
our data exhibit a behavior consistent with
the power law of QDS and pQCD.
Figure 4 (bottom) shows values for the ``reduced" deuteron
form factor\cite{BR75} $f_d(Q^2) \equiv F_d(Q^2)/F_N^2(Q^2/4)$ where the
two powers of the nucleon form factor $F_N(Q^2)=(1+Q^2/0.71)^{-2}$
remove in a minimal and approximate way the effects of nucleon 
compositeness\cite{BR75}.  Our $f_d(Q^2)$ data appear to follow, 
for $Q^2>2$~(GeV/c)$^2$, the asymptotic $Q^2$ prediction of pQCD\cite{BR83}:
$f_d(Q^2) \sim [\alpha_{s}(Q^{2})/Q^{2}]
[\ln(Q^{2}/\Lambda^{2})]^{-\Gamma}$.  Here $\Gamma=-(2C_F/5\beta)$, where
$C_F=(n_c^2-1)/2n_c$, $\beta=11-(2/3)n_f$, with $n_c=3$ and $n_f=2$
being the numbers of QCD colors and effective flavors.
Although several authors have questioned the validity of QDS and pQCD 
at the momentum transfers of this experiment\cite{IS99},\cite{FA95}, similar
scaling behavior has been reported in deuteron
photodisintegration cross sections at moderate photon
energies\cite{BO98}. 
 
In summary, we have measured the elastic structure function $A(Q^2)$
of the deuteron up to large momentum transfers.  The results have
clarified inconsistencies in previous data sets at low $Q^2$.  The
high luminosity and unique capabilities of the JLab facilities
enabled measurements of record low cross sections 
(the average cross section for $Q^2=6$~(GeV/c)$^2$ is 
$\sim 2 \times 10^{-41}$ cm$^2$/sr) that allowed extraction 
of values of $A(Q^2)$ lower by one order of magnitude than achieved at SLAC.  
The precision of our data will provide severe
constraints on theoretical calculations of the 
electromagnetic structure of the two-body nuclear system.
Calculations based on the relativistic impulse approximation
augmented by meson-exhange currents are consistent with the present data.  
The results are also
indicative of a scaling behavior consistent with predictions
of dimensional quark scaling and perturbative QCD.  Future
measurements, at higher $Q^2$, of $A(Q^2)$ and $B(Q^2)$ as well as 
of the form factors of the helium isotopes would be critical 
for testing the validity of the apparent scaling behavior. 

We acknowledge the outstanding support of the staff
of the Accelerator and Physics Divisions of JLab
that made this experiment possible.  We are grateful to the
authors of Refs. \cite{OR95},\cite{HU90} and \cite{WI95}
for kindly providing their theoretical calculations, and to F.~Gross
for valuable discussions.
This work was supported in
part by the U.S. Department of Energy and National Science
Foundation, the Kent State University Research Council, the Italian
Institute for Nuclear Research, the French Atomic Energy Commission
and National Center of Scientific Research, the Natural Sciences and
Engineering Research Council of Canada and the Fund for Scientific 
Research-Flanders of Belgium.

\end{document}